\documentclass[]{JHEP3}
\pdfoutput=1

\usepackage{graphicx} \usepackage{dcolumn}
\usepackage{bm}

\newcommand{\ft}[2]{{\textstyle\frac{#1}{#2}}}

\newcommand{\be}{\begin{equation}} \newcommand{\ee}{\end{equation}}

\newcommand{\bea}{\begin{eqnarray}} \newcommand{\eea}{\end{eqnarray}}

\def\ft12{\frac{1 2}}
\def\ft14{\frac{1 4}}

\def\atz0{|_{z=z_0}}
\def\Atz0{\bigg|_{z=z_0}}

\title{F-term uplifting and moduli stabilization consistent with K\"ahler
invariance}

\author{A.~Ach\'ucarro \\ Lorentz Institute of Theoretical Physics,
Leiden University \\ 2333 RA Leiden, The Netherlands \\ E-mail:
\email{achucar@lorentz.leidenuniv.nl} \\ {\rm and} \\  Department of
Theoretical Physics, University of the Basque Country  UPV-EHU \\
48080 Bilbao, Spain} \author{K. Sousa \\   
Lorentz Institute of Theoretical Physics,
Leiden University \\ 2333 RA Leiden, The Netherlands \\ E-mail:
\email{kepa@lorentz.leidenuniv.nl} }

\abstract {An important ingredient in the construction of
  phenomenologically viable superstring models is the uplifting of
  Anti-de Sitter supersymmetric critical points in the moduli sector
  to metastable Minkowski or de Sitter vacua with broken
  supersymmetry.  In
  all cases described so far, uplifting results in a
  displacement of the potential minimum away from the critical point
  and, if the uplifting is large, can lead to the disappearance of the minimum
  altogether.  We propose a variant of F-term uplifting
  which exactly preserves supersymmetric critical points and shift
  symmetries at tree level.  In spite of a direct coupling, the moduli
  do not contribute to supersymmetry breaking.  We analyse the
  stability of the critical points in a toy one-modulus sector before
  and after uplifting,
  and find a simple stability condition depending solely on the amount
  of uplifting and not on the details of the uplifting sector.  There
  is a region of parameter space, corresponding to the uplifting of
  local AdS {\em maxima} --or, more importantly, local minima of the 
  K\"ahler function--  where the critical points are stable for
  {\em any} amount of uplifting. On the other hand, uplifting to
  (non-supersymmetric) Minkowski space is special in that all SUSY
  critical points, that is, for {\em all} possible compactifications,
  become stable or neutrally stable.}


\preprint{}

\begin{document}
\bibliographystyle{JHEP}
\section{Introduction}
\label{introduction}

The uplifting of supersymmetric critical points from Anti-de Sitter to
Minkowski or de Sitter vacua is a crucial but still not completely
understood element in the standard KKLT scenario of moduli
stabilization in type IIB string theory \cite{Kachru:2003aw}. The
dilaton and complex structure moduli are stabilized by fluxes while
other non-perturbative effects stabilize the remaining K\"ahler moduli
at constant values that preserve supersymmetry -- leading to a
cosmological AdS vacuum with negative potential energy.  In the
original model, uplifting to a positive value of the potential is
achieved by anti-D3 branes which break supersymmetry explicitly.
Subsequent work has concentrated mainly on D-term
\cite{Burgess:2003ic,Villadoro:2005yq,Achucarro:2006zf,Choi:2006bh,Dudas:2006vc,Haack:2006cy,Burgess:2006cb,Cremades:2007ig,deCarlos:2007dp,dudasFI}
and F-term uplifting
\cite{Saltman:2004sn,GomezReino:2006dk,Lebedev:2006qq,Lebedev:2006qc,Dine:2006ii,Kitano:2006wm,Dudas:2006gr,Kallosh:2006dv,
Abe:2006xp,Abe:2007yb,Lalak:2007qd,Brax:2007xq}
as interesting alternatives where there is an explicit supergravity
description and supersymmetry is only broken spontaneously, which
gives better calculational control. Uplifting by K\"ahler corrections
has also been considered
\cite{Balasubramanian:2004uy,Westphal:2005yz}.

In the case of F-term uplifting, one possible strategy is to combine
the moduli with another sector whose SUSY breaking properties and
phenomenology are known in isolation (Polonyi model \cite{Polonyi}, O'Raifertaigh \cite{O'Raifeartaigh:1975pr},
ISS \cite{Intriligator:2007py}) and hope -or, rather, check- that the interaction with the moduli
will respect the basic features of both sectors. There are examples in
the literature both with and without direct couplings in the
superpotential between the modulus and the SUSY breaking sector.

Intuitively, one would expect compactification to be a high energy
phenomenon, possibly near the Planck scale, and therefore decoupled
from the low energy effective action describing our current Universe.
This is certainly our experience. In spite of a plethora of very
precise cosmological and accelerator data, we still see no evidence of
extra dimensions.  Upcoming experiments such as the Large Hadron
Collider at CERN or the Planck mission may change this picture but in
any case the effect is so small that it still makes sense to look for
a general framework in which at least some of the moduli are
completely stabilized and as decoupled as possible from phenomena far
below the compactification scale. Since gravity couples to all fields
and supersymmetry restricts the form of the interactions, this task
has proved somewhat tricky in supergravity.

The broader question we revisit here is how to couple two
supergravity sectors in such a way that they interact {\em as little
as possible}. We must stress that this is not a well-defined
condition, as the answer depends strongly on what properties we wish
to preserve. From the point of view of low-energy phenomenology,
requiring gravitational-strength couplings may be sufficient; however,
at higher energies this condition can become difficult to check
explicitly when there are moduli or inflatons involved with
near-Planckian vacuum expectation values. Instead, in these regimes,
supersymmetry seems a much more powerful guiding principle and one
that has been very successful in other contexts. It also facilitates
comparison with string theory, where the supersymmetry of some
configurations can be determined explicitly without reference to N=1
supergravity.

A particularly challenging decoupling problem is encountered when
trying to construct stringy models of slow roll inflation (for a recent review see
\cite{Burgess:2007pz} and references therein).  In general it is
impossible to know if a given field is a good candidate for an
inflaton until the complete potential is known because the fields
always evolve in the steepest direction of the potential. Even if the
slow-roll conditions are satisfied for a given field one must make
sure that all other fields are properly stabilized. To make matters
worse, in the standard KKLT and racetrack \cite{Kachru:2003sx,
BlancoPillado:2004ns,BlancoPillado:2006he}
scenarios, inflation can easily destabilize the moduli, leading to
decompactification.  It is therefore important to understand what kind
of supergravity lagrangians have interactions between the inflaton and
the moduli such that, on the one hand, the slow-roll conditions are
not spoilt and, on the other, no modulus becomes unstable during
inflation. Shift symmetries are sometimes invoked to solve the first
problem, for instance in relation to BPS configurations of D3-D7
branes \cite{Hsu:2003cy,Firouzjahi:2003zy,KalloshHsu}\footnote{Shift symmetries have also been used to solve the $\eta$ problem in more general contexts \cite{Kawasaki:2000yn}.}. The second
condition, in the absence of finetuning, seems to lead to the generic
constraint that the scale of inflation has to be below the gravitino
mass \cite{Kallosh:2004yh,Kallosh:2007ig}.  Therefore it is important to find
phenomenologically viable models with this property or else find ways
of evading the constraint.

In this paper we propose a way of coupling supergravity sectors that
preserves some of their supersymmetry properties such as
supersymmetric critical points and shift symmetries, at tree level. A
very important clue comes from K\"ahler invariance because the
properties we wish to preserve are invariant under K\"ahler
transformations. We will require the total action of the coupled
sectors to be invariant under K\"ahler transformations of the
individual sectors.  This limits the validity of our approach since
each sector must have an independent description in terms of N=1
SUGRA, with a non-zero superpotential, and we assume all fields are
K\"ahler invariant.  The assumption of non-zero superpotentials is
reasonable since our world, the visible sector, is
described by a near--Minkowski vacuum with broken supersymmetry, and
it also holds generically for the moduli sector \cite{Kachru:2003aw}.
Moreover, $D-$term uplifting is not possible without $F-$term
uplifting as well \cite{Choi:2005ge,deAlwis:2005tf}, so we concentrate
here on the simplest case of uncharged chiral superfields, say
$\{z^\alpha\}$ for the supersymmetric ("moduli'') sector and
$\{\phi^i\}$ for the SUSY-breaking sector responsible for the
uplifting.  In that case the D-terms are zero and the two sectors are
fully described by K\"ahler functions\footnote{We use units with $M_P
= 1$ throughout} $G^{(1)} (z, \bar z) = K^{(1)}(z, \bar z) +
\ln|W^{(1)}(z)|^2$ and $G^{(2)}(\phi, \bar\phi) = K^{(2)}(\phi, \bar
\phi) + \ln|W^{(2)}(\phi)|^2$. $K^{(1)}$ and $K^{(2)}$ are, as usual,
the K\"ahler potentials that determine the scalar manifold metric and
$W^{(1)}$ and $W^{(2)}$ the holomorphic superpotentials.  The
condition of K\"ahler invariance then tells us that, if the K\"ahler
potential $K$ of the coupled system is of the form \be K =
f~(~K^{(1)}, K^{(2)}~) \qquad {\rm or, \ more \ generally, } \qquad
F~(~K,K^{(1)},K^{(2)}) = 0 \ee for some function $F$, the K\"ahler
function of the coupled system $G$ must be of the form \be G =
f~(~G^{(1)}, G^{(2)}~) \qquad {\rm or, \ more \ generally, } \qquad
F~(~G,G^{(1)},G^{(2)}~) = 0 \ \ .  \ee In the particular case where
the kinetic terms of the two sectors are decoupled and the K\"ahler
potential is separable, $$ K (z^\alpha, \bar z^{\bar \alpha}, \phi^i,
\bar \phi^{\bar i}) = K^{(1)}(z^\alpha, \bar z^{\bar\alpha}) +
K^{(2)}(\phi^i, \bar \phi^{\bar i}) \ \ ,$$ this prescription leads
uniquely to the ansatz \be G(z^\alpha, \bar z^{\bar \alpha},\phi^i,
\bar \phi^{\bar i}) = G^{(1)}(z^\alpha, \bar z^{\bar \alpha}) +
G^{(2)} (\phi^i, \bar \phi^{\bar i}) \ ,
\label{addG}\ee that is, to the {\em product} (as opposed to the {\em
  sum}) of superpotentials
\begin{eqnarray}
  K(z^\alpha, \bar z^{\bar \alpha},\phi^i, \bar \phi^{\bar i}) &=& K^{(1)}(z^\alpha, \bar z^{\bar \alpha}) + K^{(2)}(\phi^i, \bar \phi^{\bar i}) \label{ansatzK}\\
  W(z^{\alpha}, \phi^i) &=& W^{(1)}(z^\alpha) W^{(2)}(\phi^i). 
\label{ansatz}
\end{eqnarray}
This ansatz is not new.  Binetruy \emph{et al.} \cite{Binetruy:2004hh} discuss it as a sufficient condition for integrating out  
 heavy chiral multiplets in a supersymmetric way.
Later Hsu \emph{et al.} \cite{Hsu:2003cy} used this ansatz to characterize  an effective SUGRA theory describing D3-D7 brane inflation in a type IIB string compactification.

By contrast, the usual ansatz invoked for gravitational strength
couplings,
\begin{eqnarray}
  K(z^\alpha, \bar z^{\bar \alpha},\phi^i, \bar \phi^{\bar i}) &=& K^{(1)}(z^\alpha, \bar z^{\bar \alpha}) + K^{(2)}(\phi^i, \bar \phi^{\bar i})               \\
  W(z^{\alpha}, \phi^i) &=& W^{(1)}(z^\alpha) + W^{(2)}(\phi^i). 
\label{addKW} 
\end{eqnarray}
suffers from an ambiguity in the case where the superpotentials are nonzero, since it depends on the K\"ahler gauge
chosen in each sector before combining them. A K\"ahler transformation
of each sector separately, $K^{(I)} \to K^{(I)} + 2 \;
\mathrm{Re}f^{(I)}$, and $W^{(I)} \to W^{(I)}e^{-f^{(I)}}$, $I=1,2$,
leads to \bea
K =  K^{(1)}(z,\bar z) + K^{(2)}(\phi,\bar \phi) &\to& K+ 2 \mathrm{Re} (f^{(1)}(z) +f^{(2)}(\phi) )\\
W = W^{(1)}(z) + W^{(2)}(\phi) &\to& W^{(1)}(z)e^{-f^{(1)}(z)} +
W^{(2)}(\phi)e^{-f^{(2)}(\phi)} \eea which is equivalent to \bea
K &=&  K^{(1)}(z,\bar z) + K^{(2)}(\phi,\bar \phi) \\
W &=& W^{(1)}(z)e^{f^{(2)}(\phi)} + W^{(2)}(\phi)e^{f^{(1)}(z)}\;, \eea
a completely different final theory with direct couplings between the
two sectors. The relation between the ansatz (\ref{addKW}) and
gravitational strength couplings is therefore more subtle than is
usually assumed. 

As we mentioned before, the ansatz that we propose to
couple sectors (\ref{addG}) exactly preserves supersymmetric critical
points, in contrast with the standard ansatz (\ref{addKW}), which
generically leads to a shift of these points. If we take $z^\alpha_0$
and $\phi^i_0$ to be supersymmetric critical points of the $z$ and $\phi$
sectors respectively: \be [\partial_\alpha W^{(1)}+ \partial_\alpha
K^{(1)} W^{(1)} ]_{z^\alpha_0} = 0 \qquad [\partial_i W^{(2)}+
\partial_i K^{(2)} W^{(2)}]_{\phi^i_0}=0, \ee the field configuration
$(z_0^{\alpha},\phi^i_0)$ in general will not be a SUSY critical point
of the total theory defined by (\ref{addKW}): \be [\partial_\alpha W
+\partial_\alpha K W]_{z^\alpha_0,\phi^i_0}=\partial_\alpha K^{(1)}
W^{(2)} |_{z^\alpha_0,\phi^i_0} \qquad [\partial_i W +\partial_i K
W]_{z^\alpha_0,\phi^i_0} = \partial_i K^{(2)} W^{(1)}
|_{z^\alpha_0,\phi^i_0} \ee In order to preserve the supersymmetric
critical points additional conditions must be imposed, either the
superpotentials of the individual sectors vanish 
at the critical point
$W^{(1)}|_{z^\alpha_0}=W^{(2)}|_{\phi^i_0}=0$ or the first derivatives
of the K\"ahler potential $\partial_\alpha
K^{(1)}|_{z^{\alpha}_0}=\partial_i K^{(2)}|_{\phi^i_0}=0$ are   zero at
the critical point, or some other suitable combination that makes both F-terms zero. The moduli sectors appearing in the KKLT framework
generically lead to a SUSY critical point where the superpotential does
not vanish, but the second condition can be satisfied provided an
explicit K\"ahler transformation is performed before the
superpotentials are added.\\
The paper is organized as follows. In section \ref{reviewSUGRA} we
will introduce our notation while reviewing some basic features of
$\mathcal{N}=1$ SUGRA actions. In section \ref{Ftermkahler} we study the
coupling of two sectors following the ansatz (\ref{addG}) and its
basic, model-independent properties.  We are interested in
applications to F-term uplifting so we consider a supersymmetric
sector described by an arbitrary K\"ahler function admitting one or
more critical points, which are also critical points of the potential.
The uplifting sector is also arbitrary except for the requirement that
it must have a SUSY breaking, 
Minkowski or de Sitter
vacuum or plateau at tree level -the latter, e.g. from a shift
symmetry-.  In section \ref{toymodel} we look at the stability of the
uplifted moduli in the simplest possible case: a toy model consisting
of one supersymmetric ``modulus'' field coupled to a supersymmetry
breaking ``uplifting'' sector consisting of neutral scalar fields. 
We conclude with a summary of the main results in section \ref{discussion}.

\section{Review of $\mathcal{N}=1$ supergravity}
\label{reviewSUGRA}

We start with a quick review of the relevant SUGRA formulae to fix our
notation. We take $M_{\rm Planck} = 1$. Consider an $\mathcal {N}= 1$
SUGRA sector that we will call the {\em supersymmetric sector}
consisting of neutral chiral superfields $\{ z^\alpha\} $. It is
described by a K\"ahler potential $K(z, \bar z)$ and a superpotential
$W(z)$. The kinetic terms are
$$\int d^4x \sqrt{-g} \; K_{\alpha \bar \beta} \; \partial_\mu z^\alpha \partial_\nu
\bar z^{\bar \beta} g^{\mu\nu}.$$
We will use the standard notation denoting derivatives by subscripts:
\be
\partial_{\alpha} K \equiv K_{\alpha} \qquad \partial_{\bar \beta} K
\equiv K_{\bar \beta} \quad \partial_{\alpha \bar \beta}K \equiv
K_{\alpha \bar \beta} \quad \mathrm{etc} \ldots , \ee and the indices
being raised and lowered with the K\"ahler metric $K_{\alpha \bar
  \beta}$ and its inverse $K^{\alpha \bar \beta} = K^{-1}_{\alpha \bar
  \beta }$.  Since the fields are uncharged and there are no gauge
fields, the D-terms are zero and the potential is given by \be V = e^K
\bigl[ K^{\alpha \bar \beta} (\partial_\alpha W +
\partial_\alpha K \; W) (\partial_{\bar \beta} \bar W + \partial_\beta
K \; \bar W) - 3 |W|^2\bigl] \ \ .\ee In what follows we shall omit
the superscripts $\alpha$ and $i$ of the fields.  The action and the
supersymmetry transformations are invariant under K\"ahler
transformations, \bea K (z, \bar z) &\to & K (z, \bar z) + f(z) + \bar
f (\bar z) \\ W (z) &\to & W (z) e^{-f(z)} \eea where $f(z)$ is an
arbitrary holomorphic function. If $W\neq 0$, both can be expressed in
terms of the {\em K\"ahler function}, \be G (z, \bar z) = K(z,\bar z)
+ \ln |W(z)|^2, \ee which is invariant under K\"ahler transformations.
In particular, since $G_{\alpha \bar \beta} = K_{\alpha \bar \beta}$,
the kinetic term $T$ and potential $V$ can be written as: \be T =
G_{\alpha \bar \beta} \partial_\mu z^\alpha \partial_\nu \bar z^{\bar
  \beta} g^{\mu\nu} \qquad V = e^G \bigl[ G^{\alpha \bar \beta}
G_{\alpha} G_{\bar \beta} - 3 \bigl] \label{potential}\ee The points,
$z=z_0$, where the F-terms vanish,
$$ D_\alpha W \atz0 = \partial_\alpha W \atz0+ \partial_\alpha K
\atz0 W \atz0 = 0 \qquad \Leftrightarrow \qquad \partial_\alpha G \atz0 = 0 $$
are called SUSY critical points. They are automatically critical points of $V$ because
\be \partial_\gamma V \atz0 = \biggl[ G_{\gamma} V + e^G \partial_\gamma (
G^{\alpha \bar \beta} G_{\alpha}
G_{\bar \beta} ) \biggr]  \atz0  = 0
\ee
Unlike in global SUSY, where supersymmetric critical points are always
absolute minima of $V$, the critical points in SUGRA may be local
minima, maxima or saddle points. In SUGRA, supersymmetric critical points are 
always\footnote{The case $e^{G(z_0)}=0$,
  which corresponds with a Minkowski spacetime is excluded by the
  condition $W\neq 0$} AdS since $V(z_0) = -3 e^{G
  (z_0)} $.  This means  that local maxima or saddle points
are not necessarily unstable before uplifting
\cite{Breitenlohner:1982bm}.  However, after uplifting to Minkowski or
dS, only local minima are stable.  In a Minkowski background, the
gravitino mass is $m_{3/2}^2 = e^G$.

In the next section we consider uplifting to positive $V$ by coupling
the $\{z^\alpha\}$ fields to another set of fields $\{ \phi^i\}$ that
we shall name the {\em uplifting} sector, also composed of neutral
chiral superfields. Ultimately, 
 the visible sector must also be included but this is beyond the scope of this
paper. Here we are interested in the effect of uplifting on the
moduli.

\section{F-term uplifting consistent with K\"ahler invariance}
\label{Ftermkahler}

We consider the coupling of two sectors with neutral chiral
superfields $\xi^I = z^\alpha, \phi^i$. 
We assume that each sector has a SUGRA description with a well-defined
K\"ahler function (a non-zero superpotential). If the sectors are sufficiently decoupled we expect the kinetic terms to add at tree level without interaction, so we take 
\be K =
K^{(1)}(z, \bar z) + K^{(2)}(\phi, \bar \phi)
\label{addK}
\ee
 which makes the K\"ahler metric block diagonal, and thus the kinetic terms decouple:
\be 
K_{I\bar J} \partial_\mu \xi^I\partial^{\mu} \bar\xi^{\bar J} = K^{(1)}_{
  \alpha \bar\beta} (z, \bar z) \partial_\mu z^\alpha \partial^{\mu} {\bar
  z}^{\bar \beta} + K^{(2)}_{ i \bar j} (\phi, \bar \phi)\partial_\mu
\phi^i \partial^\mu {\bar \phi}^{\bar j} 
\ee

As explained in the introduction, the ansatz (\ref{addK}) plus the
condition of invariance under K\"ahler transformations of the individual sectors:
\bea 
K^{(1)} (z, \bar z) \to K^{(1)} (z, \bar z) +
f^{(1)}(z) + \bar f^{(1)} (\bar z) &\qquad&  W^{(1)} (z) \to W^{(1)} (z) e^{-f^{(1)}(z)} \nonumber \\
K^{(2)} (\phi, \bar \phi) \to K^{(2)} (\phi, \bar \phi) +
f^{(2)}(\phi) + \bar f^{(2)} (\bar \phi) &\qquad&  W^{(2)} (\phi) \to W^{(2)} (\phi) e^{-f^{(2)}(\phi)}
\label{Ktrans}
\eea
for arbitrary $f^{(1)}(z)$ and $f^{(2)}(\phi)$,  forces us to add the full K\"ahler functions:
\be 
G (z, \bar z, \phi, \bar \phi) \equiv K + \ln |W|^2 = A(z, \bar
z) + B(\phi, \bar \phi) \label{coupling}
\ee 
where $A$ and $B$ are the corresponding
K\"ahler functions for both sectors. In our previous notation, $A
\equiv G^{(1)} \equiv K^{(1)} + \ln |W^{(1)}|^2, \ \ B \equiv G^{(2)} = K^{(2)} + \ln
|W^{(2)}|^2$. The potential becomes
\be 
V = e^G \bigl[ G^{I\bar J} G_I G_{\bar J} - 3 \bigl] = 
e^{A + B}  \bigl[ A^{\alpha \bar\beta} A_{\alpha}
 A_{\bar\beta} + B^{i\bar j}  B_i  B_{\bar j}- 3 \bigr]
\label{totalV1}
\ee 
Note that the first term in the square bracket is a function of $(z, \bar
z)$ only, and the uplifting is provided by the second term, $ B^{i\bar j}
B_i B_{\bar j}$, which is a function of $(\phi, \bar \phi)$ alone. The
exponential outside the bracket provides a direct coupling between
the two sectors.
Alternatively, we can write
\be
V = e^{B} V_A (z) + e^A V_B(\phi) + 3 e^{A+B} 
\label{potentialAB}
\ee
where
 $V_A (z) = e^A \bigl[ A^{\alpha \bar\beta}  A_{\alpha}
 A_{\bar\beta} - 3]$ would be the potential calculated for
 the $z$ sector alone and similarly  for $V_B(\phi)$.\\
 
\subsection{Critical points and stability; SUSY breaking}
As we pointed out  in the introduction, coupling the uplifting sector to the supersymmetric sector according to the ansatz (\ref{coupling}) respects the SUSY properties of the individual sectors. In particular the supersymmetric critical points of the $z$-sector are still critical points of the full potential. 
To see this suppose $z=z_0$ is a SUSY critical of the $z$-sector, $\partial_\alpha A (z_0)=0$,
then from (\ref{potentialAB}) we can see that $z_0$ also satisfies the necessary condition to be a critical point of the full potential:
\be
V_\alpha(z_0) = [ e^B V_{A\;\alpha}+A_\alpha \; e^A V_B +  3  A_\alpha \; e^{A+B}]_{z=z_0} =0.
\label{firstDer}
\ee and furthermore the $F$-terms for $z$ vanish in the full model:
\be|F_z|^2 = e^G G^{\alpha \bar \beta} G_{\alpha} G_{\bar
  \beta}|_{z_0} =0,\quad \textrm{since we have} \quad G_{\alpha}(z_0)
= A_{\alpha}(z_0) =0, \ee which means that the moduli sector does not
contribute to SUSY breaking at tree level.\\

For $z=z_0$ to really correspond to a critical point of the full potential we have to find a configuration $\phi=\phi_0$ so that the criticality condition for the uplifting sector $\partial_i V(z_0,\phi_0)=0$ is also satisfied. Using that $V_A \atz0 = -3 e^{A(z_0)}$ we find that the full potential evaluated at the point $z=z_0$ is given by the expression
\be
V\atz0 = e^{A(z_0)} V_B(\phi)\; , 
\label{Vz0}
\ee
which differs only from the original potential of the uplifting sector
by an overall factor $e^{A(z_0)}$. Therefore in order to be at an
extremum (local minimum) of the full potential we just have to
fix the uplifting
sector at any extremum (local minimum) of $V_B$, which we denote by $\phi=\phi_0$.\\

Now we turn to the issue of stability of the critical point
$(z_0,\phi_0)$ with $z_0$ being a supersymmetric critical point of the
$z-$sector. An interesting feature of this model is that it is enough
to analyze the stability of the critical point along the $\phi^i$ and
$z^\alpha$ directions separately. Indeed, the the mass matrix has a
block diagonal form, i.e. $V_{\alpha i}(z_0,\phi_0) = V_{\alpha \bar
  i}(z_0,\phi_0)=0$, and therefore we just have to check whether the
eigenvalues of the matrices \be \left(\begin{array}{cc}
    V_{\alpha \bar \beta}(z_0,\phi_0) & V_{\alpha \beta}(z_0,\phi_0) \\
    V_{\bar \alpha \bar \beta}(z_0,\phi_0) & V_{ \bar \alpha
      \beta}(z_0,\phi_0)
\end{array}\right)\quad \textrm{and} \quad
\left(\begin{array}{cc}
V_{ i \bar j}(z_0,\phi_0) & V_{i j}(z_0,\phi_0) \\ 
V_{\bar i  \bar j}(z_0,\phi_0) & V_{\bar i j}(z_0,\phi_0)
\end{array}\right)
\ee 
are all positive.
The cross terms, $V_{\alpha i}$ and $V_{\alpha \bar i}$, can be calculated taking the derivatives of (\ref{firstDer}) w.r.t.  $\phi^i$ and $\bar \phi^{\bar i}$. Then, using that $A_z$ and $V_{A \; z}$ are zero at $z=z_0$, it is easy to check that they will all vanish once evaluated at the critical point:
\begin{eqnarray}
V_{\alpha \, i}|_{z=z_0} &=& [B_i e^B V_{A \; \alpha} + A_\alpha e^A V_{B \; i} + 3 e^{A+B} A_\alpha B_i]_{z=z_0} =0\; ,\nonumber\\
V_{\alpha \, \bar i}|_{z=z_0} &=& [B_{\bar i} e^B V_{A \; \alpha} + A_\alpha e^A V_{B \; \bar i} + 3 e^{A+B} A_\alpha B_{\bar i}]_{z=z_0} =0.
\label{crossterms}
\end{eqnarray} 
Before we continue discussing the stability of the critical point let
us make some remarks about the uplifting of the SUSY critical point of
the $z-$sector, $z_0$. For later convenience let us introduce the
notation \be b(\phi)= B^{i\bar j} B_i B_{\bar j},
 \label{b}
 \ee
so that the potential of the uplifting sector alone, $V_B$, reads
\be 
V_B(\phi) = e^{B(\phi)} [b(\phi) - 3]. 
\label{VBb} 
\ee The quantity $b$ is related to the F-terms calculated from the
$\phi$ sector alone $|F_\phi|^2 = e^B \; b$, and therefore to the SUSY
breaking scale $M_s$, since
$|F_\phi| = M_s^2$. \\
In view of equation (\ref{Vz0}), which gives the vacuum expectation
value of the full potential with $z$ fixed at $z_0$, it is now clear
that in order to uplift the SUSY critical point to Minkowski or de
Sitter, we have to stabilize the $\phi-$sector
at Minkowski of de Sitter vacuum of $V_B$, $\phi_0$, so that $V_B|_{\phi=\phi_0}=0$ or $V_B|_{\phi=\phi_0}>0$ respectively. Thus for uplifting to Minkowski we need the $\phi-$sector to be stabilized at a point $\phi_0$ with $b(\phi_0)=3$, while for uplifting to de Sitter  we have to require $b(\phi_0)>3$. \\

As we argued above in order to analyze the stability of the critical
point $(z_0,\phi_0)$ it is enough to study the stability along the
$z^\alpha$
and $\phi^i$ directions separately, since the cross terms of the mass
matrix vanish (\ref{crossterms}). Therefore in order to analyze the
stability along the $z^{\alpha}$ directions it is enough to study the
potential evaluated at $\phi=\phi_0$, which reads:
\be
V|_{\phi=\phi_0}= e^{B(\phi_0)}[V_A(z)+e^{A(z)} b(\phi_0)].
\label{totalV4}
\ee
From this equation it is clear that the result of the stability
analysis
will depend on the
uplifting sector only through the value of $b(\phi_0)$.
A remarkable property of our model is that \emph{all} SUSY critical
points
of the $z-$sector are either stable or marginally stable in the
$z^\alpha$
directions after the uplifting to Minkowski vacuum $(b=3)$.
To see this we set $b(\phi_0)=3$ in the previous equation
(\ref{totalV4}),
then the full potential evaluated at the point $\phi_0$ reads
\be
V|_{\phi=\phi_0} = e^{A(z)+B(\phi_0)} A^{\alpha \bar \beta } A_\alpha A_{\bar \beta}  \ge 0 \quad \textrm{for all $z$.}
\label{positive}
\ee Since, by assumption, $V(z_0,\phi_0)=0$, the condition
(\ref{positive}) implies that no fluctuation of the fields on the
$z-$sector can decrease the energy, and therefore the point
$(z_0,\phi_0)$ is either a local minimum or a plateau along the
$z^\alpha$ directions. A similar result was found in
\cite{BlancoPillado:2005fn}. Here Blanco-Pillado \emph{et al.} argued
that SUSY vacua with vanishing cosmological constant are automatically
stable, up to flat directions. Note that such minima necessarily have
a vanishing superpotential, while in our case we are assuming that the
superpotential does not  vanish at the critical point. The main
difference is that the Minkowski critical point in our model is \emph{not
  supersymmetric}, but the coupling to the SUSY breaking sector using (\ref{addG}) respects 
the supersymmetric character of the z-sector enough to ensure the 
stability of the critical point along the $z^\alpha$ directions.\\

In general we cannot make a similar statement for uplifting to de
Sitter critical points, and the stability will depend
on the masses of the $z^\alpha$ fields before the uplifting and the value of $b$. However the analysis of the stability simplifies for large values of the uplifting parameter $b$. With the total potential written as in (\ref{totalV4}) we can see that for high values of $b(\phi_0)$ the second term dominates, and therefore the minima of $e^A$ are the ones that will survive the uplifting. Moreover, the higher the value of $b$ the higher the masses of the $z^\alpha$ fields will be after the uplifting. Note also that minima of $e^A$ are not necessarily minima of $V_A$. In the one-modulus  example described in section \ref{toymodel} the minima of $e^A$ correspond to either local maxima or  saddle points of the potential $V_A$.\\

We will now comment on the stability of the critical point
$(z_0,\phi_0)$ along the $\phi^i$ directions.  Since the stability
analysis along the $\phi^i$ directions is decoupled from the one along
the $z^\alpha$ directions, it is enough to consider the potential once
evaluated at $z=z_0$ (\ref{Vz0}).  In view of this equation we can
conclude that the minima of the combined potential coincide with the
minima of the potential of the uplifting sector before the
combination, $V_B(\phi)$, and in general it has to be checked case by
case.  In the special case of uplifting to Minkowski we just argued
that the critical point is stable, or marginally stable along
$z^\alpha$ directions, so it is evident that the problem of uplifting
the SUSY vacua of the $z-$sector to Minkowski has now reduced to
finding the stable Minkowski minima of the uplifting sector. The
conditions for the existence of SUSY-breaking Minkowski vacua have
been extensively analyzed by Gomez-Reino and Scrucca
\cite{GomezReino:2006dk,GomezReino:2006wv,GomezReino:2007qi} as well as in \cite{BlancoPillado:2005fn}. \\

Finally, in these Minkowski backgrounds, the gravitino mass after
uplifting is related to the gravitino mass of the uplifting sector
alone by \be m^2_{3/2} = e^{A(z_0)} m^2_{3/2, \phi}\ee which
is a special case of the more general relation \be e^{G (z_0, \phi_0)} =
e^{A(z_0)} e^{B(\phi_0)} \ee

\subsection{Supersymmetric critical points; BPS configurations} 

As we discussed in the previous subsection, when the $z^\alpha$ fields
are stabilized at a SUSY critical point of the $z-$sector,
$\partial_\alpha A|_{z=z_0}=0$, there is no contribution from this
sector to SUSY breaking in the full theory at tree level, i.e. the
$F-$terms associated to these fields vanish.  Therefore, for the
complete theory to be at a SUSY critical point we just have to impose
the additional condition that the F-terms for $\phi$ also vanish:
$|F_{\phi}|^2 = e^G G^{i \bar j} G_i G_{\bar j}|_{\phi=\phi_0} =0$,
which is satisfied if and only if the $\phi^i$ fields are stabilized
at SUSY critical point of the $\phi-$sector, $G_i|_{\phi=\phi_0}=B_i
(\phi_0)=0$. In other words, after fixing the $z$-fields at the SUSY
critical point of the $z$-sector, the remaining effective
theory for the $\phi$ fields gives the correct information about the critical points {\em of the full theory}.\\

This is closely related to the idea of integrating out
heavy chiral multiplets \emph{supersymmetrically}, which was first
considered in \cite{Binetruy:2004hh} (see also \cite{deAlwis:2005tg})
and it is no accident that they found the same ansatz (\ref{addG}).
Suppose that the $z^\alpha$ fields belong to the heavy chiral
multiplets we want to integrate out.  Then, when the energy scale
under consideration is much lower than the masses of the $z-$sector we
can fix these fields to their v.e.v.'s to a good approximation. If the
$z-$sector is stabilized at a SUSY critical point, the effective low
energy theory for the $\phi-$sector has unbroken $\mathcal{N}=1$ local
supersymmetry and is described by the K\"ahler function: \be
G_{eff}(\phi) \equiv G|_{z=z_0} = A(z_0) + B(\phi), \ee which
according to our previous argument would give the correct SUSY
critical points in the full theory, since $\partial_i G_{eff} =
\partial_i B$.\\

Other supersymmetry properties are also correctly inferred from the
``effective'' $\phi$-theory, for instance BPS configurations of the
$\phi$-sector are also BPS in the coupled theory since the $z$ fields
do not contribute to SUSY breaking.\\

\bigskip\noindent
\subsection{Shift symmetries}

Whenever the K\"ahler function has a shift symmetry\footnote{Note that
  we are considering shift symmetries of the full K\"ahler function
  $G$, and not just of the K\"ahler potential $K$, so that the
  statement is consistent with K\"ahler invariance.}, as $G(z+\bar
z)$, or $G(\phi+\bar \phi)$ there is a flat direction in the
potential. For example if we assume \be (\partial_z -\partial_{\bar
  z}) G = 0 \quad \textrm{we have} \quad (\partial_z - \bar
\partial_{\bar z}) V = 0 \quad , \qquad V = V(z+ \bar z).  \ee The
ansatz (\ref{coupling}) ensures that the shift symmetries of $A$ or
$B$ are also shift symmetries of full K\"ahler function $G$. Then if
one of the two sectors
has a flat direction in the potential which is related to a shift symmetry in its K\"ahler function, the same flat direction will survive in the full potential.  This statement is still true for an arbitrary number of coupled sectors.\\

We can find an example of this situation in \cite{Hsu:2003cy}.
 Here Hsu \emph{et al.} give an effective SUGRA description
of D3-D7 brane inflation in a type IIB string compactification. The
D3-D7 configuration is BPS, resulting in a supersymmetric flat
direction of the potential, which corresponds to the distance between
the D7 and the D3 branes. Such a flat direction was implemented by
introducing a K\"ahler function with a shift symmetry: \be G =
-3\ln(\rho + \bar\rho) - {(S-\bar S)^2\over 2} + \ln |W_{\rm KKLT}
(\rho)|^2 \nonumber \ee where $\rho$ is the volume modulus, $W_{\rm
KKLT} = W_0 + Ae^{-a\rho}$ is the KKLT potential and $S$ is a modulus
describing the relative distance between a probe D7 brane and a heavy
stack of D3 branes. The scalar potential derived from this K\"ahler
function is independent of $\mathrm{Re}(S)$ as a consequence of the
shift symmetry in $G$.\\

In order to be able to use the shift symmetry as an inflationary
trajectory, first we have to uplift it to de Sitter. In ref
\cite{Hsu:2003cy} the uplifting was achieved with D-terms. If we want to do
the uplifting  with F-terms while preserving the shift
symmetry at tree level we can find two possibilities. We can add a new
sector to the theory that is responsible for the uplifting and couple
it using the ansatz (\ref{addG}), or we can add a term $\Delta G
(S-\bar S)$ to the K\"ahler function such that, the $S-$sector is the
one playing the role of the uplifting sector, which has been coupled
using (\ref{addG}).  We have studied the first possibility in a toy
model with a single modulus in the $z-$sector. The results can be
found in subsection \ref{example2}.\\ 

Ref. \cite{Casas:1999xj} also describes a way to obtain exactly flat inflationary trajectories at tree level where the vacuum energy is also $F-$term dominated, but unlike in our case, the flat direction is not associated with a shift symmetry of the K\"ahler function.

\section{Moduli stabilization in a toy model}
\label{toymodel}

In this section we will study in detail the stability properties of a
simple model consisting of one modulus field $z$ and one or more uplifting
fields $\phi^i$, with K\"ahler function \be G = A(z, \bar z) + B(\phi^i,
\bar \phi^{\bar i}) \ .\label{addGAB}\ee

We start by calculating the mass spectrum of the modulus at the
critical point before we couple the uplifting sector. We then
calculate the stability after uplifting to dS or Minkowski space. The
conclusions are summarized in figure \ref{fig1}.

\subsection{Stability of the critical point before uplifting}

In order to find the mass spectrum we expand the potential around the
supersymmetric critical point $z_0$, $z = z_0 + \hat z$: \be V =
V(z_0) + {\frac 1 2} V_{zz}(z_0) \; {\hat z}^2 + {\frac 1 2} V_{\bar z
  \bar z}(z_0) \; {\hat {\bar z}}^2 + V_{z \bar z}(z_0)\; {\hat z} {\hat
      {\bar z}}+ \ldots, \ee
The diagonalization of the mass matrix gives us the spectrum of masses
squared: \be m_\pm^2 = ( V_{z \bar z}(z_0) \pm |V_{zz}(z_0)|)/ A_{z \bar z}(z_0) \label{mass2}.
\label{stability}
\ee The condition for a local minimum is therefore $V_{z \bar z}(z_0) >
|V_{zz}(z_0)| >0$. Now we will write this condition in terms of the
K\"ahler function $A$. Using \ref{potential}, with $G$ replaced by $A$, we find for the first
derivative of the potential:
\begin{equation}
 V_z = A_z V + e^{A} (A^{z\bar z}_z A_z A_{\bar z}+A^{z \bar z} A_{zz} A_{\bar z} + 
A^{z \bar z} A_z A_{z \bar z})
\end{equation}
Then the second derivatives evaluated at the critical point $z=z_0$ read:
\bea
V_{zz}(z_0) &=& - A_{zz}(z_0)e^{A(z_0)} \\
 V_{z\bar z}(z_0) &=& e^{A(z_0)}[A^{z \bar z} |A_{zz}|^2-2 A_{z \bar z} ]_{z=z_0}
\label{derivatives2}
\eea Here we used the assumption that we are expanding around a
critical point and thus $A_z(z_0)=A_{\bar z}(z_0)=0$. Defining \be x\equiv
\bigg|{A_{zz}\over A_{z \bar z}}\Atz0 \ ,\label{defx}\ee we find that before uplifting
\be m_{\pm}^2 = e^{A(z_0)} (|x|^2 -2 \pm |x|)\ee which gives a
characterization of the critical points in terms of $|x|$\footnote{Note that the stability condition is invariant under $x \to e^{\mathrm{i} \theta} \, x $, which is related
 to the $\mathrm{U}(1)$ symmetry $z \to e^{-\mathrm{i}\frac{\theta}{2}}\, z$.}
:
\bea |x| > &2& \qquad\qquad {\rm local \ AdS \ minimum} \\
\nonumber 1< |x| < &2& \qquad\qquad {\rm AdS \ saddle \ point} \\
\nonumber |x| < &1& \qquad\qquad {\rm local \ AdS \ maximum} \\
\nonumber \eea 
Local maxima in AdS are not necessarily unstable
\cite{Breitenlohner:1982bm} but such stability information is not necessary for the
present calculation.

\subsection {Stability after uplifting.}
\begin{figure}
\centering \includegraphics[width=0.4\textwidth]{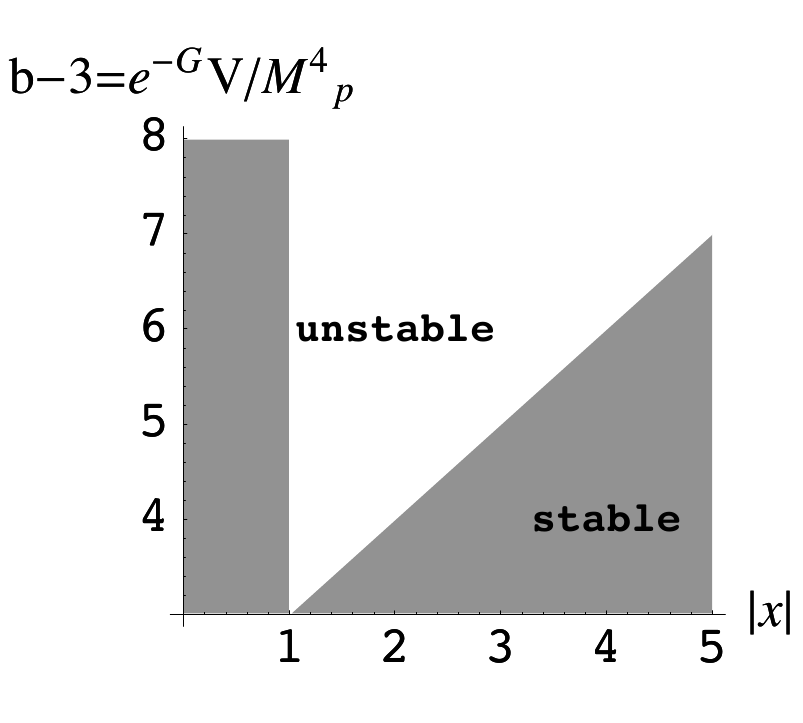}
\caption{Stability of critical points after uplifting to
Minkowski ($b=3$) or de Sitter ($b>3$) in the toy model described in
the text. The shaded areas indicate stability along the moduli $z$
directions.  The vertical axis shows the quantity $b-3=e^{-G} V/M_p^4$
evaluated at the critical point, which represents the amount of
uplifting.  The horizontal axis shows the value of the quantity $|x|=
|G_{zz}/G_{z\bar z}|\atz0$ at the critical point.  Local minima before
uplifting $(|x|>2)$ become unstable for sufficiently large
upliftings. By contrast, local maxima before uplifting $(|x|<1)$
become more stable with increasing uplifting.}
\label{fig1}

\end{figure}

Take now the K\"ahler function to be of the form (\ref{addGAB}). Then,
as it was discussed in section \ref{Ftermkahler}, after coupling the
uplifting sector $z_0$ remains a critical point of the full potential.
Moreover the mass matrix has a block diagonal form, $V_{zi}(z_0)=V_{z
  \bar i}(z_0)=0$, and therefore the stability properties of the
supersymmetric sector $A$ can be studied by just considering the
derivatives of the potential w.r.t. $z$ and $\bar z$, and the
resulting stability condition for the field $z$ is again of the form
(\ref{stability}). We just have to calculate $V_{zz}$ and $V_{z \bar
  z}$. From (\ref{potentialAB}) we obtain:
\begin{eqnarray}
V_{zz}|_{z=z_0} &=& [e^B V_{A \; zz} + A_{zz} e^A V_B + 3 A_{zz} e^{A+B}]_{z=z_0}  \\
V_{z\bar z}|_{z=z_0} &=& [e^B V_{A \; z\bar z} + A_{z\bar z} e^A V_B + 3 A_{z\bar z} e^{A+B}]_{z=z_0} .
\end{eqnarray}
We can recast these equations in a more compact form using
(\ref{defx}), and using the abbreviation (\ref{b})
$b \equiv B^{i \bar j} B_i B_{\bar j}$, which is only a function of the uplifting
sector:
\begin{equation}
V_{zz}|_{z=z_0} = e^{A+B}|_{z=z_0} (b-1) \; x  \qquad V_{z\bar z}|_{z=z_0} = e^{A+B}|_{z=z_0} (|x|^2+b-2)
\end{equation}
Here, as in the previous section $|x|=|A_{zz}/A_{z \bar z}|_{z=z_0}$. Finally we can write the spectrum of masses squared around the critical point:
\begin{equation}
  m^2_{\pm}=e^{A+B}|_{z=z_0} \biggl[ (|x|^2+b-2) \pm |(b-1) x| \biggr] = 
  e^{A+B}|_{z=z_0} \left[ \left(|x| \pm {\frac 1 2} (b-1) \right)^2 -{\frac 1 4}(b-3)^2  \right]
\end{equation}
To simplify the mass formula we assumed that $b>1$. In the opposite case, $b<1$, the masses $m_+^2$ and $m_-^2$ are exchanged. The stability condition for the field in the supersymmetric sector after uplifting reads:
\begin{equation}
  m_{\pm}^2 = e^{A+B}|_{z=z_0}  \biggl[ \left(|x| \pm {\frac 1 2}(b-1) \right)^2 - {\frac 1 4}(b-3)^2 \biggr] \geq  0
\end{equation}
The solutions to these inequalities in the case of
uplifting to Minkowski or de Sitter, $b \geq 3$, are presented in
fig.(\ref{fig1}).  We list here some interesting properties:
\begin{itemize}
\item Notice that the stability properties do not depend on the details of the uplifting sector, just on the 
amount of uplifting $b$. This actually fits in the intuition of weakly coupled systems.
\item All critical points $z_0$ that were local
  minima before the uplifting $(b=0, \ |x|>2)$ remain stable for a
  certain amount of uplifting, and then all became unstable. As an
  example, the minimum found in the original KKLT paper \cite{Kachru:2003aw} had
  $|x| \sim  25$.
\item Critical points that are local maxima with $|x|<1$ before uplifting, $b=0$, become stable for $b=3$, and  remain stable for arbitrarily higher values of $b$. 
These points correspond to 
local minima of $e^{A/2}$.\end{itemize}
In the case $x = 0$, which corresponds with having no ${\hat z}^2$
terms in $A (z,\bar z)$, the two masses are equal,  $m_\pm^2 =(b-2)
e^{A+B}|_{z_0}$, and both positive for $b> 2$. Points with $|x|=1$
have one of the masses equal to zero  for any uplifting.

Uplifting to \emph{non-supersymmetric} Minkowski vacua has a special
property.  If $b = 3$ the mass squared 
\be 
(m^2\pm)_{Minkowski} = 
m_{3/2}^2(|x|\pm 1)^2  
\ee 
is positive definite for {\em any} choice of $A(z)$
(any value of $|x|$). Here we have used that in Minkowski vacuum
the gravitino mass is given by $m_{3/2}^2 = e^{A+B}|_{z_0,\phi_0}$.
This situation is close in spirit to the global susy
case where critical points are always absolute minima. Here, Minkowski
vacua are local minima of the supersymmetric sector whenever $|x| \neq
1$ or have a zero mode when $|x| = 1$. In the next section we give an
explicit example of this latter case based on a shift symmetry.
 \subsection{A simple example: an uplifted flat direction in dS
    protected by shift symmetry} 
\label{example2}
 We will now consider the case where the K\"ahler function of 
the supersymmetric sector $A(z,\bar z)$ has a shift symmetry, we will take 
it to be of the form $A=A(z+\bar z)$. Given the shift symmetry,
$\partial_z$ and $\partial_{\bar z}$ are interchangeable when they act
on $A$ or $V$ so, in particular we have
\be 
A_z=A_{\bar z},\quad  A_{zz} = A_{z \bar z} \quad \textrm{and} \quad V_{zz} = V_{z \bar z}.
\ee
Suppose now that   $A(z,\bar z)$ has a SUSY critical
  point\footnote {The KKLT superpotential for the volume modulus with
    $W_0 = 0$ is of this form, $W = A e^{-az}$ which gives $G =
    -3\log(z+ \bar z) -a(z+\bar z) + const. $ but its SUSY critical
    point is unphysical since it has negative $z+\bar z$ at $z_0$, $A_z(z_0) =0$}. For this critical  point  we have 
      $|x| =|A_{zz}/A_{z \bar z}|=1$. Before
uplifting there is one flat direction (zero mass) and one
``tachyonic'' direction with negative mass squared
\bea
m_-^2 &=& 0 \\ 
m_+^2 &=& 2 V_{z \bar z}/A_{z \bar z} \atz0= -2 e^A \atz0 < 0  .\eea The zero
mode reflects the fact that the potential does not depend on $Im z$,
and the Re z direction is always a local maximum since $A_{zz} = A_{z
  \bar z} > 0$, although not necessarily unstable since we are in AdS.
 After uplifting,
the mass squared becomes positive while the flat direction remains
\bea
m_-^2 &=& 0 \\
m_+^2 &=& 2 V_{z\bar z}/A_{z \bar z} \atz0 = e^{A+B}|_{z=z_0}(b-1),
\eea so in this case it seems we can have positive mass squared
whenever $ b = B^{i\bar j} B_i B_{\bar j} \geq 1$, and in particular
whenever $b\geq 3$.
We note, however, that these results are only
in our toy model but whether they generalize to the case with several
moduli remains to be seen.

For $b>3$ this simple model has a de Sitter, exactly flat $z$
direction protected by the shift symmetry. Note that this ``inflaton
trench'' is an F-term-uplifted AdS ``ridge'' (a line of local maxima),
in contrast with the one proposed in \cite{Hsu:2003cy}, which was an
AdS ``trench'' uplifted by D-terms. Its viability as an inflationary
trajectory depends on whether the quantum corrections
(from couplings to other fields) will tilt the flat direction to the
required level. Alternatively, a soft breaking of the shift symmetry
can be introduced.  A graceful exit from inflation requires a more
complicated scenario. But the point we want to emphasize is
that there is no $\eta$ problem.

\section{Summary and discussion}
\label{discussion}

Motivated by the KKLT uplifting problem, we have investigated a class
of models where the K\"ahler potential and the superpotential are of
the form:
\begin{eqnarray}
  K(z^\alpha, \bar z^{\bar \alpha},\phi^i, \bar \phi^i) &=& K^{(1)}(z^\alpha, \bar z^{\bar \alpha}) + K^{(2)}(\phi^i, \bar \phi^i) \nonumber \\
  W(\phi^i,z^{\alpha}) &=& W^{(1)}(z^\alpha) W^{(2)}(\phi^i) \nonumber,
\end{eqnarray}
or, equivalently, where the full K\"ahler function is of the form \be
G = G^{(1)}(z^\alpha, \bar z^{\bar \alpha}) +
G^{(2)} (\phi^i, \bar \phi^i) \ .  \nonumber \ee
We have shown that these models have a number of interesting general
properties:
\begin{itemize}
\item If $z^\alpha=z_0^{\alpha}$ is a SUSY critical point in the model
  defined by $G^{(1)} (z^\alpha, \bar z^{\bar \alpha})$, that is, if
  $({\partial G^{(1)} / \partial z^\alpha}) |_{z^\alpha=z^\alpha_0}=0$,
  this sector will not contribute to SUSY breaking in the full model:
  $F_z \propto ({\partial G / \partial z^\alpha})
  |_{z^\alpha=z^\alpha_0} = 0$.  Moreover $z^\alpha=z^\alpha_0$ is
  automatically a critical point of the combined (uplifted) potential
  in the $z$-direction.
\item The stability of uplifted SUSY critical points of the
  $z-$sector, can be analyzed independently in the $z^\alpha$ and
  $\phi^i$ directions, since the crossed second derivatives of the
  combined potential vanish at this point: $\partial_\alpha \partial_i
  V|_{z=z_0} = \partial_{\bar \alpha} \partial_i V|_{z=z_0}=0$.
\item Local minima of the $\phi$-potential when the uplifting sector
  is considered alone --the model defined by $G^{(2)} (\phi^i, \bar
  \phi^{\bar i})$--, \emph{always} remain local minima in the
  $\phi$-directions after the uplifting of the SUSY critical point of
  the $z-$sector, $z_0^\alpha$. Moreover if the $\phi-$sector is
  stabilized at a Minkowski or de Sitter vacuum, $z_0^\alpha$ will be
  uplifted to Minkowski and de Sitter respectively.
\item When a supersymmetric critical point is uplifted to Minkowski
  the critical point becomes automatically stable or flat along the
  $z^\alpha$ directions. A similar result was obtained in
  \cite{BlancoPillado:2005fn}, where it was proven that all SUSY
  Minkowski critical points are stable. However our result describes
  the uplifting of SUSY critical points to \emph{non supersymmetric}
  Minkowski vacua.
\item When a supersymmetric critical point is uplifted to de Sitter,
  for sufficiently large cosmological constant the local minima of
  $G^{(1)}(z_\alpha, \bar z_\alpha)$ are always local minima of the
  combined potential (after uplifting) along the
  $z^\alpha$-directions. Moreover, this local minimum becomes more
  stable with increasing value of the cosmological constant. Note that
  local minima of $G^{(1)}$ are always extrema of the moduli potential
  before uplifting, but not necessarily local minima. 
\item Shift symmetries of the individual sectors survive after the uplifting, 
becoming interesting candidates for inflationary trajectories.
\end{itemize}
We have studied in detail a toy model with a single
field in the supersymmetric sector, where we have analyzed the stability of the $z-$sector
 "before", and "after" the uplifting.
We have confirmed that uplifting to Minkowski space is special in that all SUSY
  critical points (irrespective of the choice of $G^{(1)}(z, \bar z)$) become  stable or
  neutrally stable.  Indeed, after uplifting to Minkowski, the moduli
masses are given by 
\be m_\pm^2 = m^2_{3/2} (|x| \pm 1)^2 \;, \quad \textrm{with} \quad |x|= |G^{(1)}_{z z}/G^{(1)}_{z \bar z}|_{z=z_0} = |G_{z z}/G_{z \bar z}|_{z=z_0}  .\ee 
Note that if $|x|>2 $ the
masses of the scalars in the supersymmetric sector are larger than the gravitino mass, for example, in the case of the KKLT model, $|x| \sim 25$, they are considerably larger. In general for $|x|<1$, $m_\pm$ are of the order of the gravitino mass, except when the value of $x$ is very close to 1, because in this case $m_-$ becomes significantly lower than $m_{3/2}$. 
The case $|x|<1$ is interesting because an uplifted critical point is
stable for an
arbitrary amount of uplifting.  These critical points are precisely  the minima of $G^{(1)}(z, \bar z)$, which in this toy model  correspond to local AdS maxima  of the scalar potential before uplifting.\\
    
Finally, we have shown that performing K\"ahler transformations before
coupling two sectors gravitationally \be K = K^{(1)}+K^{(2)} \qquad W
= W^{(1)}+W^{(2)}\;, \nonumber \ee
may lead  to direct couplings, and therefore the choice of K\"ahler gauge 
plays an important role  in the applicability of this prescription.\\
An important consideration is whether string theory contains
sectors that are coupled in the way described in this paper. We think
it may be possible to find such couplings in certain $\mathcal{N}=2$
compactifications where the presence of fluxes breaks supergravity
down to $\mathcal{N}=1$. $\mathcal{N}=2$ supergravity requires that
the kinetic terms of the scalars of vector and hypermultiplets appear
totally decoupled from each other, and although the scalar manifold in
general gets distorted during the SSB, there are known cases where
this decoupling prevails
\cite{Louis:2002vy,Fre:1996js,Ferrara:2002bt,Andrianopoli:2002vq}.
This leads to effective $\mathcal{N}=1$ theories with a K\"ahler
potential of the form (\ref{ansatzK}).  Moreover it is known that if
the effective $\mathcal{N}=1$ resulting from the SSB is consistent
with a truncation of $\mathcal{N}=2$ there are situations where the
superpotential has a product structure \be W =W^{(1)}(vector) \;
W^{(2)}(hyper) \ee where $W^{(1)}$ and $W^{(2)}$ depend only on the
scalars that came from the $\mathcal{N}=2$ vector and hypermultiplets
respectively \cite{Andrianopoli:2001zh,Andrianopoli:2001gm}, which is
precisely the kind of structure that we have studied in this paper.

\acknowledgments
                                                                               
We thank J.A. Casas, S. Davis, M. Esole, S. Hardeman, R. Jeannerot, M.
Postma, F.  Quevedo, K. Schalm and S. Vandoren for very useful
discussions.  This work was supported in part by the Netherlands
Organisation for Scientific Research (N.W.O.)  under the VICI
Programme, by the Basque Government grant BFI04.203 and by the
Spanish Ministry of Education through project FPA2005-04823.

\bibliography{Ftermuplifting}

\end{document}